\documentclass[12pt,reqno]{amsart}
\usepackage{amsmath,amsfonts,amssymb,amscd,amsthm,amsbsy, color, siunitx}
\usepackage[dvips]{graphicx}
\usepackage{comment}
\usepackage[format=plain,justification=raggedright,singlelinecheck=false]{caption}
\usepackage{appendix}

\usepackage{wrapfig}
\usepackage{fancybox}
\usepackage[maxfloats=100]{morefloats}

\usepackage{multimedia}
\usepackage{graphicx,epsfig}

\numberwithin{equation}{section}

\newtheorem{meta-thm}[theorem]{Meta-Theorem}

\newcommand\beq[1]{ \begin{equation}\label{#1} }
\newcommand{\eeq}{ \end{equation} }

\newcommand\beqa[1]{ \begin{eqnarray} \label{#1}}

\newcommand{\eeqa}{ \end{eqnarray} }
\newcommand{\beqano}{ \begin{eqnarray*} }
\newcommand{\eeqano}{ \end{eqnarray*} }
\newcommand\equ[1]{{\rm (\ref{#1})}}

\def\T{{\mathcal T}}

\begin{document}

\title[Da Keplero alla materia oscura]
{Dalle leggi di Keplero alla materia oscura attraverso la storia
di donne scienziate}


\author[A. Celletti]{Alessandra Celletti}

\address{
Department of Mathematics, University of Rome Tor Vergata, Via
della Ricerca Scientifica 1, 00133 Rome (Italy)}

\email{celletti@mat.uniroma2.it}


\baselineskip=18pt              

\maketitle


\begin{flushright}
\sl La scienza progredisce meglio

quando le osservazioni ci costringono

a cambiare i nostri preconcetti.

\vskip.1in

Vera Rubin \rm
\end{flushright}

\vskip.2in

\begin{flushright}
\sl Dedicato alle donne di questo pianeta,

in particolare afghane e iraniane,

a cui \`e negata l'istruzione e quindi la libert\`a.\rm
\end{flushright}

\vglue1cm

\bf Sommario: \rm L'universo \`e composto da materia ordinaria (ad
esempio, stelle, pianeti, noi stessi), da materia oscura ed
energia oscura. La percentuale stimata di materia oscura \`e di
circa il $27\%$ dell'intero universo, mentre il $5\%$ \`e la
materia ordinaria e la parte rimanente \`e l'energia oscura.
Questo articolo introduce gli ingredienti fondamentali che hanno
portato ad ipotizzare l'esistenza della materia oscura e a
studiarne le possibili propriet\`a. A tale risultato hanno
contribuito, direttamente o indirettamente, numerose matematiche e
astronome, che ricordiamo per dare loro il giusto credito.

\vskip.1in

\bf Abstract: \rm The universe is composed of ordinary matter (for
example, stars, planets, ourselves), dark matter and dark energy.
The estimated percentage of dark matter is about $27\%$ of the
entire universe, while $5\%$ is ordinary matter and the remaining
part is dark energy. This article introduces the fundamental
ingredients that led to conjecture the existence of dark matter
and to study its properties. To this result have contributed,
directly or indirectly, numerous female mathematicians and
astronomers, who we quote to give them the right credit.

\vglue1cm

\section{Introduzione}\label{sec:intro}

\begin{figure}[htpb]
\vglue6cm \hglue-3cm
 \includegraphics[scale=0.1]{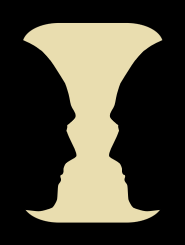}
 \caption{Un vaso bianco e due profili contrapposti di volti in nero (immagine creata da Edgar Rubin,
 fonte Wikipedia)}
 \label{fig:1}
\end{figure}

Da bambina mi capit\`o di sfogliare la rivista "Le Scienze" (n.
43, marzo 1972), che conteneva un articolo intitolato
``Multistabilit\`a della percezione" ad opera di Fred Attneave,
esperto di percezione visiva. L'articolo conteneva immagini che
possono essere interpretate in maniera diversa a seconda di come
il lettore le percepisce. Un esempio  \`e mostrato nella
Figura~\ref{fig:1}, che riporta un'immagine creata intorno al 1915
dallo psicologo danese Edgar Rubin\footnote{Non sono al corrente
di una relazione di parentela tra Edgar Rubin e Vera Rubin.}.
L'effetto visivo presenta al lettore due differenti
interpretazioni, a seconda della percezione che stimola: un vaso
bianco al centro o due profili di volti in nero contrapposti.
Ognuna delle due percezioni \`e consistente con l'immagine
retinica a seconda del modo in cui viene attribuito il contorno,
ovvero come delimitazione di una delle due forme, mentre l'altra
viene percepita come sfondo. Ma si tratta appunto di una
percezione, perch\'e l'immagine \`e unica e non c'\`e confine tra
le due interpretazioni.

Allo stesso modo il contenuto di questo articolo potr\`a essere
interpretato in maniera diversa a seconda della percezione del
lettore: da una parte illustrer\`o la storia di alcune donne
scienziate che hanno fornito contributi importanti alla matematica
e all'astronomia; dall'altra cercher\`o di spiegare gli
ingredienti scientifici che hanno portato ad ipotizzare
l'esistenza della materia oscura, che costituisce una parte
considerevole dell'universo, circa il $27\%$. Il lettore potr\`a
dunque percepire questo lavoro come un'esposizione su questioni di
genere o piuttosto come un articolo su un argomento scientifico
per spiegare le motivazioni alla base della materia oscura. Il mio
obiettivo sar\`a piuttosto di cercare di far percepire l'articolo
come un'unica storia, senza confine tra scienza e scienziati, tra
donne e uomini, tra matematica e astronomia, perch\`e di fatto la
storia \`e una sola, al di l\`a della percezione individuale.

Spiegare il percorso che ha portato alla congettura dell'esistenza
della materia oscura richiede diversi ingredienti scientifici. Il
primo \`e costituito dalle leggi di Keplero
(Sezione~\ref{sec:Keplero}) che descrivono molto spesso una buona
approssimazione delle traiettorie dei corpi celesti. Tali leggi
valgono in un modello in cui si considerano solo due corpi, ad
esempio il Sole e la Terra; in tale modello approssimato si pu\`o
dimostrare che le traiettorie sono delle coniche: ellissi (se il
sistema ha energia negativa), parabole (se il sistema ha energia
nulla), iperboli (se il sistema ha energia positiva). Se si
considera un altro corpo celeste, ad esempio Giove, si passa al
problema dei tre corpi, nell'esempio precedente si dovrebbe
considerare la terna Sole-Terra-Giove; tale modello diventa assai
pi\`u complesso del problema dei due corpi: le relative
traiettorie possono essere regolari o piuttosto possono esibire
caratteristiche della dinamica del caos (Sezione~\ref{sec:caos}).
Il sistema solare nella sua interezza \`e composto da moltissimi
oggetti ed \`e quindi un sistema di incredibile complessit\`a. Le
galassie sono sistemi ancor pi\`u dinamicamente complessi, in
quanto sono formate da miliardi di stelle. Comprendere la
classificazione di stelle e galassie \`e di grande importanza per
i nostri scopi (Sezione~\ref{sec:stelle}). Per studiarne la
dinamica \`e fondamentale l'uso di computer potenti
(Sezione~\ref{sec:computer}), attraverso cui si ottengono
simulazioni sulla composizione dell'universo, inclusa la materia
oscura (Sezione~\ref{sec:darkmatter}). Allo sviluppo di tutti
questi ingredienti hanno contribuito\footnote{Per chiarire l'uso dell'asterisco qui e nel seguito, molt* scienziat* \`e da leggersi
come: molte scienziate e molti scienziati.} molt* scienziat*, tra
cui numerose scienziate di grande spessore scientifico.

\section{Keplero e le leggi di Keplero}\label{sec:Keplero}

La dinamica della Terra, dei pianeti e di tutti gli oggetti del
sistema solare \`e di grande complessit\`a, perch\`e coinvolge la
reciproca interazione gravitazionale di un grandissimo numero di
corpi celesti. Per affrontare il problema del calcolo delle
traiettorie celesti, \`e necessario partire da un'approssimazione
in cui si considerano solamente due oggetti, ad esempio il Sole e
la Terra, e si trascurano tutti gli altri corpi che compongono il
sistema solare. Questa approssimazione viene chiamata problema dei
due corpi o problema di Keplero. La sua soluzione venne infatti
elaborata da Johannes Kepler, italianizzato in Keplero, nella
forma di tre leggi che ci accingiamo a descrivere brevemente
(\cite{Alebook}). Nell'enunciare tali leggi, prenderemo come
esempio la Terra e il Sole, ma ovviamente esse si applicano ad una
qualsiasi coppia di oggetti che si attraggono secondo la legge di
Newton (ad esempio, il Sole e un asteroide, oppure un pianeta e un
suo satellite).

In linea di principio, le soluzioni del problema dei due corpi
soggetti alla reciproca attrazione gravitazionale sono ellissi,
parabole e iperboli, a seconda del valore dell'energia del
sistema; tuttavia, noi considereremo solo orbite ellittiche, come
quelle dei pianeti intorno al Sole.

\begin{enumerate}
    \item[$(i)$] La prima legge stabilisce che la Terra si muove attorno al Sole
lungo un'ellisse di cui il Sole occupa uno dei due fuochi e
l'altro fuoco \`e vuoto.
    \item[$(ii)$] La seconda legge afferma che la congiungente Sole-Terra spazza aree uguali in
tempi uguali; come conseguenza si ha che la Terra \`e pi\`u veloce
al perielio (punto di massima vicinanza al Sole lungo l'ellisse),
che all'afelio (punto di massima lontananza dal Sole).
    \item[$(iii)$] La terza legge stabilisce una relazione tra il periodo di
tempo impiegato a descrivere l'ellisse e il corrispondente
semiasse maggiore.
\end{enumerate}

La terza legge di Keplero sar\`a di fondamentale importanza per
comprendere come si \`e arrivati alla materia oscura. Prima di
illustrare gli ingredienti scientifici che ci serviranno nel resto
di questo lavoro, descriviamo un particolare aspetto del contesto
in cui viveva Keplero e la sua famiglia.

La terza legge di Keplero \`e contenuta nel trattato ``Harmonices
Mundi"; scritto nel 1619, il trattato contiene una discussione
delle analogie tra l'armonia della musica e il moto dei pianeti.
Alcuni anni pi\`u tardi Keplero revision\`o l'``Harmonices Mundi",
ma allo stesso tempo si impegn\`o in una causa completamente
trasversale rispetto ai suoi interessi scientifici: scrisse una
memoria per difendere la madre Katharina, che era stata accusata
di stregoneria. Sposata con un soldato mercenario che presto
abbandon\`o la famiglia, Katharina ebbe quattro figli; visse in
una cittadina, Leonberg in Germania, in cui il governatore locale,
Lutherus Einhorn, accus\`o 15 donne di stregoneria e giustizi\`o 8
di loro.

Erano periodi terribili; nell'arco di due secoli, tra il 1500 e il
1700, vennero giustiziate circa 50\,000 persone di cui il $75\%$
erano donne. L'accusa a Katharina venne formulata da una vicina di
casa, Ursula Reinbold: Katharina era responsabile di averle dato
una pozione di erbe che le procur\`o problemi di salute.
Probabilmente Katharina non  aveva alcuna intenzione malvagia, ma
era solamente una donna anziana, colpevole di curiosit\`a e
fantasia che la portarono ad inventare bevande mediche a base di
erbe. Il 7 Agosto 1620 venne arrestata all'et\`a di 73 anni; oltre
alle accuse della Reinbold, il giudice le contest\`o di
comportarsi in maniera anomala per i seguenti due motivi: durante
il processo non guard\`o mai i testimoni e non vers\`o neanche una
lacrima. Il figlio Johannes tent\`o di giustificare il
comportamento della madre, sostenendo che non aveva mai avuto
particolari espressioni del viso e che probabilmente \`e un fatto
naturale che le lacrime si seccano in alcune persone anziane.
Durante il  processo che dur\`o circa un anno, Katharina venne
imprigionata e incatenata. Johannes all'epoca aveva 44 anni e
rivestiva la carica di matematico imperiale di Rodolfo II; per
amore filiale, decise di assumere l'incarico di difensore della
madre.

Con lo stesso rigore con cui aveva individuato le leggi basilari
della Meccanica Celeste, Johannes smont\`o tutte le accuse contro
la madre; richiese alla corte la trascrizione di tutti gli atti
(\cite{Caspar1959}) e prepar\`o una difesa di 128 pagine
eccezionalmente efficace, in cui identific\`o tre cause principali
per la persecuzione di sua madre: la sua debole situazione sociale
in quanto vedova, il timore della gente comune verso le donne
anziane, l'eccessiva reazione del governatore contro le donne.
Attacc\`o anche l'affidabilit\`a dei testimoni, sia perch\`e erano
troppo giovani o semplicemente perch\`e erano donne (solo gli
uomini erano affidabili, mentre le donne erano tipicamente ingenue
e superstiziose). L'ordalia finale consistette nel sottomettere
Katharina ad una tortura verbale. Un giustiziere le mostr\`o gli
strumenti di tortura e se lei avesse confessato, sarebbe stata
giustiziata. Tuttavia, Katharina pronunci\`o le seguenti parole:
``Anche se mi trattatte in qualsiasi modo, e strappate una vena
dopo l'altra dal mio corpo, non saprei cosa dovrei ammettere".
Nell'ottobre 1621 venne prosciolta dall'accusa di stregoneria e
liberata, ma tanta era stata la sofferenza durante i mesi di
prigionia che mor\`i nell'aprile del 1622.

Ci\`o accadde proprio nel periodo in cui Keplero revision\`o
l'``Harmonices Mundi" e rielabor\`o la terza legge dei moti
planetari. Fedele sostenitore della teoria copernicana, Keplero
ebbe la fortuna di lavorare insieme ad un astronomo danese, Tycho
Brahe (1546-1601), che era eccezionalmente bravo nel costruire
degli strumenti che consentivano di compiere osservazioni
estremamente precise di stelle e pianeti. Sulla base dei dati
acquisiti assieme a Brahe, Keplero scopr\`i le tre leggi
$(i)-(ii)-(iii)$ che governano la dinamica di due corpi celesti
soggetti alla mutua attrazione gravitazionale. \`E interessante
sottolineare che le tre leggi vennero elaborate da Keplero senza
conoscere la legge di gravitazione di Isaac Newton (che nacque 13
anni dopo la morte di Keplero), ma solamente sulla base dei dati
sperimentali. La terza legge stabilisce che il periodo del moto
cresce con la distanza dal Sole e collega il periodo $T$ del moto
al semiasse maggiore $a$ dell'orbita ellittica secondo la formula
(in opportune unit\`a di misura):
$$
T^2 = a^3\ .
$$
Come conseguenza della terza legge di Keplero, la velocit\`a
orbitale diminuisce con la distanza dal Sole; in particolare
Mercurio orbita con una velocit\`a di circa 47.9 km/s, la Terra
circa 29.8 km/s, Saturno circa 9.6 km/s, Nettuno circa 5.4 km/s.
La formula che descrive tale conseguenza si ottiene uguagliando la
forza centripeta e la forza gravitazionale: \beq{va}
v_P(R)=\sqrt{{GM(R)}\over R}\ , \eeq dove $v_P(R)$ \`e la
velocit\`a del pianeta a distanza $R$, $G$ la costante di
gravitazione universale, $M(R)$ la massa contenuta entro la
distanza $R$ (nel sistema solare $M(R)$ \`e ben approssimata dalla
massa del Sole). La Tabella~\ref{tab:velocita} fornisce i
risultati che si ottengono applicando la formula \equ{va},
confrontati con i valori forniti dalla NASA (\cite{NASAv})
relativi alla velocit\`a media dei pianeti lungo la loro orbita
attorno al Sole.

\vskip.2in

\begin{table*}
  \centering
\caption{Velocit\`a $v_P$ dei pianeti usando la formula \equ{va}
che fornisce la velocit\`a in funzione del semiasse maggiore $a$ e
confronto con i valori $v_{NASA}$ della NASA (\cite{NASAv}).}
      \label{tab:velocita}
\begin{tabular}{|c|c|c|c|c|c|c|c|c|}
  \hline
   & Mercurio & Venere & Terra & Marte & Giove & Saturno & Urano & Nettuno \\
  \hline
  $a$ ($10^6$ km) & 57.9 & 108.2 & 149.6 & 227.9 & 778.6 & 1433.5 & 2872.5 & 4495.1 \\
  $v_P$ (km/s) & 47.9 & 35.0 & 29.8 & 24.1 & 13.0 & 9.6 & 6.8 & 5.4 \\
    $v_{NASA}$ (km/s) & 47.4 & 35.0 & 29.8 & 24.1 & 13.1 & 9.7 & 6.8 & 5.4 \\
  \hline
\end{tabular}
\end{table*}

\begin{figure}[htpb]
\vglue6cm \hglue-9cm
 \includegraphics[scale=0.1]{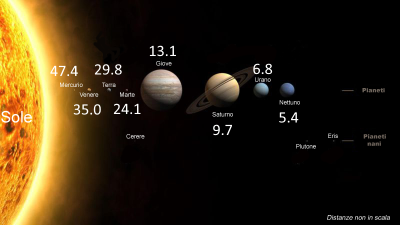}
 \caption{Il sistema solare con l'indicazione delle velocit\`a dei pianeti in km/s (adattata da Wikipedia).}
 \label{fig:speed}
\end{figure}

La Figura~\ref{fig:speed} mostra una rappresentazione grafica (non
in scala) dei pianeti del sistema solare e le loro velocit\`a
orbitali medie.

\vskip.2in

Si noti che l'accordo tra i due risultati presentati nella
Tabella~\ref{tab:velocita} \`e alla prima o seconda cifra
decimale. Naturalmente le stesse considerazioni valgono se invece
di esaminare il Sole e un pianeta si considerano un pianeta e un
satellite (naturale o artificiale) oppure due oggetti qualsiasi
che si attragono gravitazionalmente. Applicato a due stelle della
nostra o di altre galassie, il collegamento tra velocit\`a e
distanza sar\`a di fondamentale importanza per comprendere la
motivazione che port\`o a congetturare l'esistenza della materia
oscura.

\section{Alla scoperta del caos}\label{sec:caos}
Il  problema dei due corpi \`e solo una prima approssimazione
della dinamica del sistema solare che, piuttosto, risulta composto
da tantissimi oggetti: otto pianeti, numerosi satelliti, centinaia
di migliaia di asteroidi e comete. Si tratta di un sistema di
grandissima complessit\`a in cui peraltro ognuno di questi oggetti
\`e dotato di un movimento orbitale attorno al corpo centrale e
allo stesso tempo di un moto di rotazione attorno ad un asse
interno. Nel problema di Keplero, infatti, abbiamo trascurato il
movimento di rotazione, assumendo che i due oggetti siano
puntiformi.

Molto spesso l'approssimazione del problema a due corpi non \`e
sufficiente a dare una descrizione esauriente della dinamica ed
\`e piuttosto necessario considerare l'attrazione di altri corpi
del sistema solare. Procedendo per gradi, dal problema dei due
corpi si passa a quello dei tre corpi, come ad esempio il caso
della Terra soggetta alla forza gravitazionale di Sole e di Giove.
La scelta del pianeta Giove \`e dettata dal fatto che esso \`e il
corpo celeste pi\`u grande dopo il Sole ed esercita un'influenza
gravitazionale considerevole sulla Terra.

Il problema dei tre corpi \`e il modello pi\`u studiato della
Meccanica Celeste; purtroppo, come ha mostrato Poincar\'e in
\cite{Poincare99}, tale problema non ammette una soluzione
esplicita come invece accade nel problema dei due corpi. Si pu\`o
tuttavia trovare una soluzione approssimata attraverso
l'applicazione della \sl teoria delle perturbazioni. \rm
Sviluppata gi\`a a partire dal XVIII secolo, la teoria delle
perturbazioni parte dall'assunzione che si pu\`o spesso pensare il
problema dei tre corpi come un problema a due corpi, soggetto
all'ulteriore perturbazione di un terzo corpo; \`e questo il caso
di Sole-Terra-Giove, in cui il problema a due corpi Sole-Terra \`e
perturbato dall'azione gravitazionale di Giove, molto pi\`u debole
rispetto all'azione del Sole, poich\`e la massa di Giove \`e solo
un millesimo di quella del Sole. Ricordando che il problema dei
due corpi \`e integrabile, ovvero \`e risolubile esattamente (la
soluzione \`e un'ellisse), il problema dei tre corpi si pu\`o
pensare come una perturbazione di un sistema integrabile.

La teoria delle perturbazioni consente, attraverso elaborati
calcoli, di assumere come punto di partenza la soluzione del
problema dei due corpi e di trovare un'approssimazione del
problema dei tre corpi come una perturbazione dell'ellisse
Kepleriana. Il teorema che formalizza la teoria delle
perturbazioni \`e costruttivo e fornisce quindi tutti gli
ingredienti per determinare esplicitamente la soluzione
approssimata. L'applicazione pi\`u famosa della teoria della
perturbazioni \`e senz'altro la scoperta del pianeta Nettuno ad
opera di Urbain Leverrier (1811-1877). All'epoca erano state
osservate delle discrepanze tra le osservazioni astronomiche di
Urano, a quel tempo il pianeta conosciuto pi\`u lontano dal Sole,
e i calcoli matematici per determinarne la traiettoria.
Utilizzando la teoria delle perturbazioni, Leverrier determin\`o,
entro $5^o$ di precisione, la posizione presunta di un ulteriore
pianeta oltre Urano che, con la sua attrazione gravitazionale,
poteva giustificare le discrepanze tra teoria ed osservazioni.
Leverrier chiese ad un collega astronomo, Johann Galle, di
osservare il cielo nella posizione da lui predetta. Grazie ai
calcoli basati sulla teoria delle perturbazioni, nel settembre del
1846 si scopr\`i l'esistenza di Nettuno.

\vskip.1in

\begin{figure}[htpb]
\vglue7cm \hglue-5cm
  \includegraphics[scale=0.01]{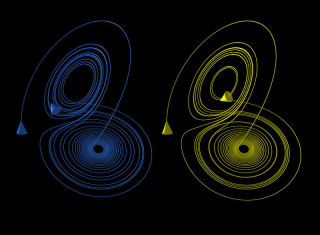}
\hglue9cm
  \includegraphics[scale=0.05]{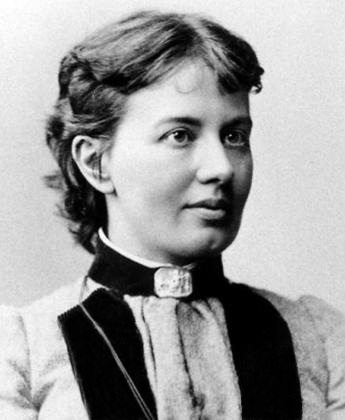}
\caption{Sinistra: la teoria del caos,
 due orbite del sistema di Lorenz con condizioni iniziali vicine
 $(0,0,1)$, $(0,0,1.00001)$ e traiettorie divergenti (fonte Wikipedia). Destra: Sofia Kovalevskaya (1850-1891) (fonte Wikipedia).}
 \label{fig:2}
\end{figure}

Un altro aspetto molto rilevante del problema dei tre corpi fu la
scoperta ad opera di Henri Poincar\'e (1854-1912) dell'esistenza
di moti caotici, ovvero caratterizzati da una dinamica che mostra
un'estrema sensibilit\`a alla scelta delle condizioni iniziali. In
altre parole, consideriamo un sistema dinamico e prendiamo due
condizioni iniziali molto vicine tra di loro. Lasciamo evolvere il
sistema secondo le equazioni che descrivono il sistema dinamico a
partire dalle due condizioni iniziali. Se la distanza tra le
corrispondenti traiettorie si mantiene limitata nel tempo, allora
il moto si definisce regolare; se, invece, si osserva una
divergenza nel tempo delle traiettorie, allora siamo in presenza
di un moto caotico. Dai tempi di Poincar\'e, la teoria del caos
torn\`o agli onori della ribalta nel 1963, quando il meteorologo
Edward Lorenz scopr\`i un sistema di tre equazioni differenziali
del primo ordine che mostrava estrema sensibilit\`a alla scelta
delle condizioni iniziali (\cite{Lorenz}). Attraverso simulazioni
al computer, Lorenz trov\`o che una variazione infinitesima delle
condizioni iniziali conduceva a due diverse soluzioni del sistema
di equazioni differenziali, l'una corrispondente a tempo sereno,
l'altra ad una tempesta. Una soluzione del sistema di Lorenz \`e
mostrata nel pannello di sinistra della Figura~\ref{fig:2}; vista
la somiglianza grafica di tale soluzione con le ali di una
farfalla, nel 1972 Lorenz intitol\`o una sua
conferenza\footnote{``Prevedibilit\`a: il battito di ali di una
farfalla in Brasile pu\`o scatenare un tornado in Texas?"}
``Predictability: Does the Flap of a Butterfly's Wings in Brazil
Set Off a Tornado in Texas?". Nasce cos\`i l'iconico ``effetto
butterfly" come sinonimo di caos.

\vskip.1in

\`E interessante osservare che ai tempi di Poincar\'e la teoria
del caos permeava l'ambiente culturale, anche a livello di
concetto letterario. L'aneddoto che segue coinvolge la matematica
Sofia Vasilevna Kovalevskaya (1850-1891), talvolta chiamata con il
diminuitivo Sonya Kovalevskaya. Allo stesso tempo in cui
Poincar\'e si dedicava al problema dei tre corpi, Sofia
Kovalevskaya studiava la rotazione di un corpo solido attorno a un
punto fisso e dimostrava che, solo in casi speciali, tale problema
\`e integrabile ovvero risolubile esattamente (\cite{Cooke}). Per
questo risultato nel 1888 vinse il Premio Bordin dell'Accademia
Francese delle Scienze. Tali soluzioni integrabili sono casi assai
rari, in coerenza con il fatto che il problema dei tre corpi non
\`e integrabile. Sofia Kovalevskaya non si occup\`o della teoria
matematica del caos, ma si avvicin\`o al caos in maniera piuttosto
singolare. Infatti, la Kovalevskaya coltivava anche una passione
per la letteratura; scrisse diverse commedie, romanzi, poemi, tra
cui la commedia ``La lotta per la felicit\`a: come \`e stato e
come sarebbe potuto essere", scritta con Anna Carlotta Leffler. La
commedia, disponibile in svedese, si compone di due atti che si
svolgono parallelamente nel tempo: entrambi gli atti partono dallo
stesso inizio e con gli stessi protagonisti, ma le storie hanno
destini diversi nei due atti. La trama della commedia coinvolge
due coppie di giovani. Nel primo atto, ``Come \`e stato", Alisa,
di nobili origini, rifiuta il povero spasimante Karl e, per un
senso del dovere, sposa il cugino Hjalmar, il quale rinuncia
all'amata Paula, sorella di Karl, per sposare la ricca Alisa.
Tutti sono infelici, Karl sposa un'altra donna e Hjalmar si
suicida. Nel secondo atto, ``Come sarebbe potuto essere", si parte
dalla stessa trama con il matrimonio tra Alisa e Hjalmar. I
protagonisti trovano per\`o il coraggio di cambiare le loro vite:
Alisa e Hjalmar divorziano, Karl e Alisa si sposano, Hjalmar si
dichiara a Paula. La commedia termina con Karl e Alisa che creano
una cooperativa nella loro fabbrica, e decidono di dividere gli
introiti con gli impiegati. Con un po' di fantasia, possiamo
interpretare la commedia come una trasposizione letteraria della
teoria del caos.

E a proposito di letteratura, visto che le poche fotografie
disponibili non ritraggono mai la Kovalevskaya ampiamente
sorridente, vale la pena di concludere con la frase che a lei
viene attribuita in un racconto del premio Nobel per la
letteratura Alice Munro (\cite{munro}): ``Sembro piuttosto seria
nelle fotografie perch\'e la gente non avrebbe fiducia in me se
sorridessi".

\section{La classificazione delle stelle}\label{sec:stelle}

Per arrivare alla materia oscura abbiamo bisogno di altri
ingredienti e, in particolare, abbiamo bisogno di approfondire la
conoscenza di stelle e galassie. Andiamo all'inizio del '900 e
cominciamo dal racconto della storia di Williamina Paton Stevens
(1857-1911). Nata in Scozia, lavor\`o come insegnante per alcuni
anni e nel 1877 spos\`o James Orr Fleming; entrambi si
trasferirono negli Stati Uniti, ma poco dopo il loro arrivo in
America, il marito abbandon\`o la moglie, proprio mentre aspettava
un bambino. La Fleming aveva necessit\`a di lavorare per mantenere
il figlio e se stessa. Non avendo alcuna preparazione
specialistica, e in particolare scientifica, accett\`o un lavoro
come governante presso la famiglia di Edward Charles Pickering
(1846-1919), un famoso astronomo di Harvard. Pickering rivest\`i
la carica di direttore dell'Osservatorio Astronomico di Harvard
nel periodo 1877-1919. I progetti scientifici di Pickering
consistevano in studi ottici e fotografici delle stelle, e
nell'analisi spettroscopica per poter fotografare, misurare e
classificare gli spettri stellari. Per raggiungere i suoi
obiettivi scientifici, aveva necessit\`a di assumere numerosi
assistenti, pi\`u economici possibile, per compiere lavori di
routine come calcolare e copiare. L'obiettivo principale era di
completare il catalogo dell'astronomo Henry Draper (1837-1882),
pioniere dell'astrofotografia. Tra i numerosi risultati
scientifici di rilievo, ricordiamo che Draper ottenne la prima
immagine dello spettro di una stella, la prima fotografia della
nebulosa di Orione, inizi\`o un catalogo stellare comprendente
dati astrometrici e spettroscopici, che raggiunse la cifra di
359\,000 stelle, anche debolmente luminose.

Nel 1879 Pickering offr\`i alla Fleming un lavoro part-time
all'Osservatorio, perch\`e si rese conto che la Fleming riusciva a
lavorare meglio di molti assistenti di genere maschile. Tra il
1885 e il 1890 assunse oltre 40 donne, che divise in
``calcolatori" e ``registratori" a seconda della mansione ad esse
assegnata; le assistenti venivano pagate 25 centesimi all'ora,
meno di una segretaria. Tra le donne che lavorarono in quello che
scherzosamente veniva chiamato l'\sl harem di Pickering \rm
troviamo scienziate di grande valore (\cite{sobel}) che hanno
scritto la storia dell'astronomia, tra cui le quattro scienziate
di cui diamo qui sotto un breve cenno biografico.

\begin{enumerate}
    \item[1)] Williamina Fleming (1857-1911) contribu\`i alla classificazione del catalogo Henry Draper, scoprendo 59 nebulose,
    310 stelle variabili, 10 novae, 94 stelle Wolf-Rayet. Fu nominata Direttore del laboratorio di fotografia,
membro della Royal Astronomical Society di Londra e dell'Astronomical Society del Messico;
    \item[2)] Annie Jump Cannon (1863-1941) entr\`o nel Wellesley College nel 1880, dove si laure\`o nel 1884.
Da giovane contrasse la scarlattina e divent\`o quasi non udente.
Nel 1896 divenne assistente di Edward Pickering; accurata e
veloce, nella sua vita riusc\`i a classificare circa 350\,000
stelle, fino a 300 all'ora. Svilupp\`o un'importante
classificazione stellare, dividendo le stelle in classi spettrali
contrassegnate con le lettere O, B, A, F, G, K, M; in alcuni libri
di testo (anche relativamente recenti) si suggeriva di ricordare
tali lettere come le iniziali della seguente frase: ``Oh Be A Fine
Girl, Kiss Me". Ottenne il dottorato honoris causa ad Oxford nel
1925, la medaglia Henri Draper nel 1931, il premio Ellen Richards
nel 1932. Fu la prima donna a diventare dirigente dell'American
Astronomical Society;
    \item[3)] Henrietta Leavitt (1868-1921) si laure\`o al Radcliffe College e dal 1902
    lavor\`o come assistente di Pickering. Interruppe il lavoro per diversi anni a causa
    di una malattia, che la rese quasi sorda. Colleg\`o i periodi di alcune stelle (le variabili
    Cefeidi) alla loro luminosit\`a intrinseca.
La relazione di Leavitt tra periodo e luminosit\`a consent\`i di
calcolare la distanza delle stelle e con tale relazione
l'astronomo Shapley determin\`o la grandezza della Via Lattea. La
misura delle distanze consent\`i di trovare stelle Cefeidi in
altre galassie, ponendo fine al grande dibattito tra gli astronomi
Shapley e Curtis, riguardante la questione dell'appartenenza di
alcune nebulose alla Via Lattea o piuttosto ad altre galassie,
separate dalla nostra galassia. Quattro anni dopo la morte della
Leavitt, il matematico G\"osta Mittag-Leffler la propose per il
premio Nobel, ma non pot\`e comunque esserle assegnato postumo;
    \item[4)] Antonia C. Maury (1866-1952), nipote di Henri Draper,
studi\`o al Vassar College come allieva di Maria Mitchell, la
prima donna astronoma professionista americana. Maury si laure\`o
nel 1887 e due anni dopo cominci\`o a lavorare come \sl computer
\rm per la classificazione di spettri stellari nell'emisfero Nord.
Rielabor\`o lo schema della Fleming e raffin\`o la classificazione
stellare introducendo un proprio sistema di classificazione. Dallo
spettro stellare ottenne la magnitudine assoluta (una quantit\`a
collegata alla luminosit\`a delle stelle) e misurando la
magnitudine apparente\footnote{La magnitudine apparente \`e un
indice di quanto ci appare luminoso un oggetto nel cielo,
mentre la magnitudine assoluta \`e la luminosit\`a di un oggetto alla distanza
fissata di 10 parsec, pari a 32.6 anni luce.} determin\`o la distanza delle stelle.
Secondo l'astronomo Ejnar Hertzsprung, il lavoro della Maury ``fu
uno dei progressi pi\`u importanti nella classificazione
stellare".
\end{enumerate}

\begin{figure}[htpb]
\vglue1cm \hglue-1cm
 \includegraphics[scale=1]{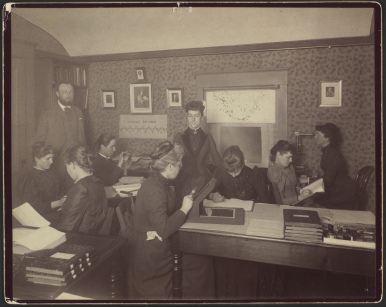}
 \caption{Il Direttore dell'Harvard College Observatory Edward Pickering (in piedi a sinistra)
 e alcune collaboratrici (si ringrazia l'Harvard University Archives).}
 \label{fig:stars}
\end{figure}

Ad ognuna di loro \`e dedicato un cratere lunare. Grazie al loro
lavoro conosciamo molte interessanti propriet\`a di stelle e
galassie. La Figura~\ref{fig:stars} mostra Edward Pickering nella
stanza in cui lavoravano alcune sue collaboratrici.

\section{La programmazione dei computer}\label{sec:computer}
Dalle leggi di Keplero siamo passati al problema dei tre corpi,
che ci ha consentito di scoprire il caos; poi siamo arrivati a
studiare le caratteristiche delle galassie, che sono sistemi a
molti corpi di grande complessit\`a, perch\`e ognuna \`e
costituita da circa 100 miliardi di stelle. Abbiamo quindi bisogno
di un altro ingrediente per poter simulare un  sistema cos\`i
complesso: un computer programmabile.

Ci spostiamo alla met\`a del XX secolo; durante la seconda guerra
mondiale, lo stato americano, carente di uomini impegnati al
fronte, assunse donne matematiche allo scopo di elaborare
complessi calcoli e precisamente per compilare le tavole
balistiche per calcolare le traiettorie dei proiettili, le
cosidette tabelle \sl bombing and firing \rm (\cite{benvenuti1}).
Si trattava di calcoli estremamente importanti per le persone in
guerra, ma assai complicati perch\`e necessitavano di risolvere a
mano delle equazioni differenziali che dipendevano da diversi
fattori, quali l'angolo di tiro, la dimensione del proiettile e
l'attrito atmosferico. La procedura consisteva nel discretizzare
le equazioni differenziali che descrivono il modello del
proiettile e nell'usare un passo di integrazione temporale molto
piccolo (0.1 secondi). Per calcolare una traiettoria di 30 secondi
erano necessarie 20 ore di calcoli a mano. Con l'utilizzo del
primo analizzatore differenziale il tempo di calcolo si ridusse a
30 minuti, ancora troppo lento. Il punto di svolta si ebbe con
l'introduzione del primo computer moderno chiamato ENIAC
(Electronic Numerical Integrator And Computer), un progetto della
US Army a Philadelphia realizzato dagli ingegneri Presper Eckert e
John Mauchly. Essi assunsero 366 matematiche per programmare la
macchina; tra di esse le matematiche Fran Bilas, Betty Jean
Jennings Bartik, Ruth Lichterman, Kay McNulty, Betty Snyder,
Marlyn Wescoff. Nella loro biografia \cite{bartik} viene detto
esplicitamente che non avevano un ufficio, non disponevano di un
manuale operativo, nessun istruttore, ma possedevano solo un
diagramma a blocchi di questa macchina parallela. In compenso
avevano un'illustre predecessore: la matematica Ada Lovelace
(1815-1852), figlia di Lord Byron e Anne Isabella Milbanke, che
scrisse il primo software di una macchina analitica ideata da
Charles Babbage.

La presentazione dell'ENIAC avvenne il 15 Febbraio 1946. La notte
prima della presentazione il computer ancora non funzionava
perfettamente, perch\'e il calcolo non si fermava quando la
traiettoria del proiettile toccava terra. Fortunatamente durante
la notte precedente la presentazione Betty Snyder realizz\`o che
era necessario cambiare una delle dieci posizioni degli oltre
3\,000 interruttori: con quest'ultimo ritocco, l'ENIAC calcol\`o
la traiettoria senza errori e pi\`u velocemente del viaggio del
proiettile.

La dimostrazione dell'ENIAC fu un vero successo; dopo la
presentazione tutti gli uomini si recarono ad una cena di
celebrazione alla Houston Hall del campus. Nella biografia scritta
dalla Bartik (\cite{bartik}) si sottolinea che le donne vennero
lasciate negli uffici, senza una parola di congratulazioni e non
vennero incluse nei festeggiamenti; tuttavia, la Bartik concluse
la descrizione di questa esperienza con la frase ``at least all of
us women could feel a quiet satisfaction in being able to say \sl
I was there"\rm\footnote{``almeno, tutte noi donne potemmo provare
soddisfazione nel poter dire ``io ero l\`i" ".}.

\begin{figure}[h]
\centering
\vglue5.7cm
\hglue-8cm
\includegraphics[scale=0.1]{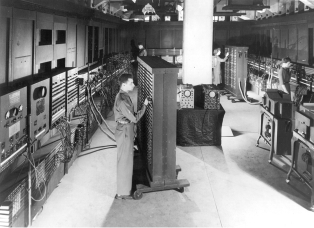}
 \caption{Un'immagine delle matematiche che lavoravano all'ENIAC
 (fonte ``U.S. Army photo, Public domain, via Wikimedia Commons").}
 \label{fig:eniac}
\end{figure}

Il loro lavoro continu\`o ad essere dimenticato fino agli anni
'80, quando una studentessa di Harvard, Kathy Kleiman, trov\`o una
fotografia che ritraeva ingegneri e progettisti insieme a due
collaboratrici donne (si veda ad esempio la
Figura~\ref{fig:eniac}). Kleiman chiese chi fossero le due donne
ad uno storico dell'informatica, professore ad Harvard, il quale
le rispose di non conoscerle e che probabilmente erano solamente
delle ``Refrigerator Ladies", ovvero donne che erano in posa
vicino alle macchine per renderle pi\`u attraenti, cos\`i come si
fa nel marketing per vendere, ad esempio, pi\`u frigoriferi
(\cite{benvenuti1}, si veda anche \cite{benvenuti2}).

Da allora, nonostante l'oblio delle prime programmatrici
dell'umanit\`a, i computer hanno avuto un inarrestabile progresso
e sono diventati uno strumento fondamentale per esplorare
l'universo e la materia oscura.

\section{La materia oscura}\label{sec:darkmatter}
Oltre alle leggi di Keplero, al caos e alle galassie, abbiamo ora
un computer che ci consentir\`a di arrivare finalmente alla
materia oscura di cui la protagonista femminile sar\`a l'astronoma
Vera Rubin (1928-2016).

\vskip.1in

Nata a Philadelphia, Vera Cooper Rubin si diplom\`o al Vassar
College; successivamente prov\`o ad entrare all'Universit\`a di
Princeton, ma non ricevette nemmeno il catalogo dei corsi offerti,
perch\'e alle donne non era consentito studiare astronomia a
Princeton. Studi\`o quindi alla Cornell University e consegu\`i il
dottorato di ricerca alla Georgetown University. Successivamente
venne invitata dall'astronomo Allan Sandage a presentare la
richiesta di utilizzo dei telescopi installati presso gli
osservatori pi\`u importanti dell'epoca, situati a Monte Wilson e
Monte Palomar. Nella sua autobiografia \cite{rubin}, si legge che
sul modulo era stampato che a causa di limitate infrastrutture non
era possibile accettare le domande presentate da donne; le
infrastrutture mancanti erano le \sl toilette \rm per signore.
Tuttavia, qualche persona particolarmente lungimirante aggiunse
sul modulo di Vera Rubin che ``di solito" non erano accettate
domande presentate da donne, lasciandole cos\`i la possibilit\`a
di inviare  la richiesta. Vera Rubin present\`o la domanda, che
venne accolta positivamente. In questo modo le venne concesso di
compiere osservazioni astronomiche assieme al collega Ken Ford,
che aveva elaborato uno spettrografo di altissima precisione.

Insieme studiarono le stelle di circa 60 galassie a spirale e
misurarono la loro velocit\`a orbitale. Come abbiamo illustrato
nella Sezione~\ref{sec:Keplero}, secondo la terza legge di Keplero
la velocit\`a delle stelle  in una galassia dovrebbe diminuire
come l'inverso della radice quadrata della distanza delle stelle
dal centro della rispettiva galassia. Al contrario, Rubin e il suo
collaboratore osservarono che le velocit\`a orbitali delle stelle
nelle galassie, invece di decrescere, rimangono costanti con la
distanza dal nucleo galattico. Una simile conclusione era gi\`a
stata formulata dall'astronomo svizzero Fritz Zwicky nel 1933, che
studi\`o le velocit\`a delle galassie (e non delle singole stelle)
nell'ammasso della Chioma e nell'ammasso della Vergine (senza
per\`o considerare il contributo del gas intergalattico formato da
materia ordinaria). La conclusione di Zwicky era basata sulla
determinazione della massa attraverso il teorema del viriale. In
un sistema autogravitante ad energia negativa (che corrisponde a
considerare orbite ellittiche nel problema dei due corpi,
piuttosto che paraboliche o iperboliche), l'energia totale \`e
somma dell'energia cinetica $\T$ e dell'energia potenziale $V$; il
teorema del viriale afferma che $\T$ e $V$ soddisfano la relazione
$$
2\langle \T\rangle_t = -\langle V\rangle_t\ ,
$$
dove $\langle \cdot\rangle_t$ denota la media temporale su un
tempo $t$ sufficientemente lungo. I termini $\T$ e $V$ dipendono
dalla massa, che \`e l'incognita dell'equazione, dalle velocit\`a,
misurabili sfruttando l'effetto Doppler, e dalle distanze relative
che, per galassie di cui \`e nota la distanza, si ottengono
misurando le separazioni angolari. Assumendo che le galassie di
uno stesso ammasso abbiano masse comparabili, il teorema del
viriale fornisce il valore della massa gravitazionale
dell'ammasso. L'eccesso di massa trovato da Zwicky utilizzando il
teorema del viriale venne attribuito alla presenza di materia non
luminosa, a complemento della materia ordinaria di cui sono
costituiti stelle e pianeti.

\begin{figure}[h]
\centering
\vglue6cm
\hglue-8cm
\includegraphics[scale=0.1]{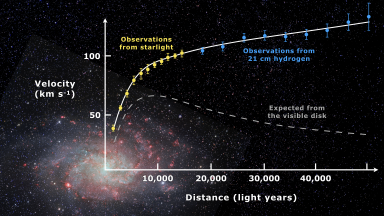}
 \caption{Curva di rotazione della galassia Messier 33 ottenuta interpolando
 i punti gialli e azzurri (con relative barre di errore) ottenuti attraverso osservazioni sperimentali;
 la previsione in base alla materia visibile \`e mostrata dalla linea tratteggiata
 (\cite{CS2000}). La discrepanza tra le due curve pu\`o essere giustificata
 dall'esistenza di un alone di materia oscura che circonda la galassia
 (fonte Wikipedia).}
 \label{fig:dark}
\end{figure}

Studiando le stelle in una galassia (invece delle galassie in un
ammasso, come fece Zwicky), Rubin e Ford (\cite{RF}) notarono che
le velocit\`a orbitali non decrescevano con la distanza secondo
quanto previsto teoricamente come mostrato nella
Figura~\ref{fig:dark}; essendo la velocit\`a costante, Rubin
congettur\`o che la diminuzione della massa brillante dal centro
della galassia doveva essere compensata dall'aumento di materia
non brillante, appunto la materia oscura o ``dark matter". In
particolare, la materia oscura non emette radiazione
elettromagnetica, ma pu\`o essere osservata solamente attraverso
la sua azione gravitazionale.

In un successivo articolo scritto da Rubin, Ford e Thonnard
(\cite{RFT}), gli autori concludono che le loro osservazioni degli
spettri e delle curve di rotazione delle galassie pongono dei
vincoli sui modelli di formazione ed evoluzione delle galassie. In
particolare, il profilo delle curve di rotazione sembra
evidenziare che la massa non \`e concentrata al centro,  ma che
una massa significativa si trova diffusa a grandi distanze dal nucleo
galattico; inoltre, la massa non converge ad un limite al confine
dell'immagine ottica e questa osservazione porta inevitabilmente a
concludere che esista un alone di materia non luminosa che si estende ben oltre la galassia
ottica\footnote{``The conclusion is inescapable that non-luminous
matter exists beyond the optical galaxy", \cite{RFT}.}. Con queste
conclusioni, l'esistenza della materia oscura ha ormai un solido fondamento
scientifico.

\begin{figure}[h]
\centering
\vglue8cm
\hglue-8cm
\includegraphics[scale=0.1]{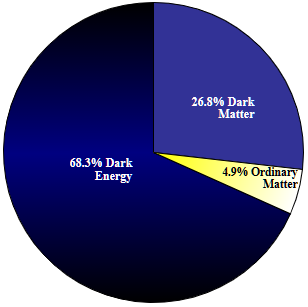}
 \caption{Composizione dell'universo in materia ordinaria, materia oscura ed energia oscura
 (fonte Wikipedia).}
 \label{fig:matter}
\end{figure}

La stima della percentuale di materia oscura che compone l'intero
universo \`e di circa il $27\%$, mentre il $5\%$ rappresenta la
materia ordinaria e la parte rimanente \`e l'energia oscura (si
veda la Figura~\ref{fig:matter}). Il termine energia ``oscura"
indica nuovamente che di tale energia si hanno solo prove
indirette, e precisamente l'espansione con accelerazione
dell'universo.

Non \`e noto di cosa sia composta la materia oscura, \`e pi\`u
semplice dire che non \`e composta da materia visibile. Possibili
spiegazioni sono al vaglio degli scienziati. Materia oscura
``ordinaria" potrebbe provenire da pianeti con una massa non
sufficiente a diventare stelle, e pertanto destinati a non
produrre energia, e quindi a non essere luminosi. Pi\`u in
generale, oltre ai pianeti, anche stelle nane brune e buchi neri
costituiscono la classe dei MACHO, ``MAssive Compact Halo
Objects", e potrebbero contribuire alla massa complessiva delle
galassie, anche se i dati del progetto EROS negli anni '90
suggeriscono che tali oggetti non sono sufficientemente numerosi
da giustificare la massa mancante.

I neutrini potrebbero contribuire alla materia oscura, ma dovrebbero
avere una massa sufficiente per giustificare l'abbondanza di
materia non visibile. I fisici provano a cercare particelle esotiche
come i \sl neutralini \rm predetti dalla teoria SUSY, SUper
SYmmetry, oppure cercano particelle ad interazione debole, che
vengono genericamente indicate con il nome WIMP, acronimo di
``Weakly Interacting Massive Particles" (\cite{bertone}).

Esiste un'alternativa alla materia oscura per giustificare la
discrepanza tra teoria e osservazioni? Una possibilit\`a \`e
rappresentata dalla teoria MOND, MOdified Newtonian Dynamics,
sviluppata dal fisico israeliano Mordehai Milgrom nel 1983;
tuttavia, tale teoria richiederebbe una modifica sostanziale della
forza di gravitazione, che a grandi distanze si dovrebbe
comportare come l'inverso della distanza, invece dell'inverso del
quadrato della distanza in accordo alla legge di gravitazione di
Newton.

\begin{figure}[h]
\centering
\vglue6cm
\hglue-6cm
\includegraphics[scale=0.1]{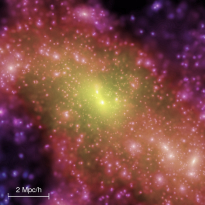}
\hglue6cm
\includegraphics[scale=0.15]{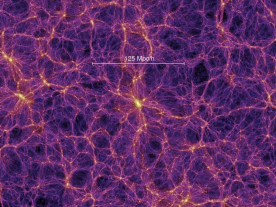}
 \caption{Sinistra: distribuzione della materia oscura nella galassia.
 Destra: distribuzione della materia oscura in un ammasso di galassie
 (fonte Springel et al., Virgo Consortium, Max-Planck-Institute for Astrophysics).}
 \label{fig:sim}
\end{figure}

Studi relativamente recenti nell'ambito del ``Millennium
Simulation Project" hanno mostrato un'interessante simulazione
della distribuzione nell'universo della materia oscura. A partire
dalla congettura di Vera Rubin e Ken Ford, la simulazione \`e
effettuata dal Max Planck Centre con un super-computer (assai
pi\`u potente di quello usato dalle ``Refrigerator Ladies"),
zoomando su un ammasso di galassie (ben studiate dalle scienziate
dell'harem di Pickering), ovvero un ammasso composto da $10^{10}$
oggetti (quindi un problema a $N$-corpi con $N=10^{10}$ di
grandissima complessit\`a, decisamente pi\`u complesso del caotico
problema dei tre corpi o del regolare problema di Keplero).

Su scale maggiori di quelle delle singole galassie, la materia oscura potrebbe disporsi in filamenti, senza
essere distribuita uniformemente e formando una sorta di
gigantesca ragnatela. Per il momento, se vogliamo averne un'idea
guardiamo la simulazione della Figura~\ref{fig:sim}, in attesa che
un'altra scienziata o un altro scienziato ci svelino di cosa \`e
composto il $27\%$ dell'universo. Non \`e poco, considerando che
si stima che la parte osservabile dell'universo abbia un raggio di
circa 46.5 miliardi di anni luce, ovvero $4.4\cdot 10^{23}$ km
ovvero 10 trilioni di giri della Terra all'equatore. Non \`e
affatto poco.

\section{Conclusioni}

Qui di seguito tre diverse possibili versioni per le conclusioni.

\vskip.1in

Versione n. 1: in questo articolo abbiamo raccontato il
contributo, diretto o indiretto, attraverso i secoli di
scienziat*, e in particolare di alcune donne (prevalentemente
matematiche e astronome), con cui si \`e arrivati alla congettura
dell'esistenza della materia oscura.

\vskip.1in

Versione n. 2: in questo articolo abbiamo descritto gli ingredienti scientifici principali,
dalle leggi di Keplero alle curve di rotazione, che, attraverso i
secoli, hanno portato ad ipotizzare l'esistenza della materia oscura.

\vskip.1in

Versione n. 3: in questo articolo abbiamo descritto gli elementi
scientifici con cui si \`e arrivati alla congettura della materia
oscura; a tale risultato hanno contribuito scienziat* di grande
valore, senza distinzione di sesso, perch\'e tutte le idee
scientifiche sono uguali ed hanno pari dignit\`a di fronte alla
Scienza\footnote{L'autrice ha preso ispirazione dalla prima parte
dell'Articolo 3 della Costituzione italiana: ``Tutti i cittadini
hanno pari dignit\`a sociale e sono eguali davanti alla legge,
senza distinzione di sesso, di razza, di lingua, di religione, di
opinioni politiche, di condizioni personali e sociali". Sarebbe
auspicabile che ognuno di noi ricordasse e mettesse in pratica
anche la seconda parte dell'Articolo 3, che afferma: ``\`E compito
della Repubblica rimuovere gli ostacoli di ordine economico e
sociale, che, limitando di fatto la libert\`a e l'eguaglianza dei
cittadini, impediscono il pieno sviluppo della persona umana e
l'effettiva partecipazione di tutti i lavoratori
all'organizzazione politica, economica e sociale del Paese".}.

\vskip.1in

Lascio al lettore la scelta della conclusione preferita, senza
dimenticare che ad un risultato cos\`i importante (ed affatto
trascurabile visto che si stima che la quantit\`a di materia
oscura sia oltre un quarto dell'universo) probabilmente non si
sarebbe arrivati se nel cammino si fossero interposti... una
toilette per signore, un frigorifero, un harem, un sorriso, o una
strega, forse, ma comunque una strega senza lacrime.

\vskip.2in

\bf Ringraziamenti. \rm Prima di scrivere questo articolo, ho
presentato il suo contenuto in alcuni eventi che hanno contribuito a
consolidare il racconto. Tra questi, la conferenza pubblica per la
EWM/EMS a Berlino nell'ambito del 7th ECM il 20  Luglio 2016,
l'intervista a RadioTre Scienza condotta magistralmente da
Rossella Panarese il 3 Gennaio 2017, la conferenza della Classe di
Scienze Fisiche, Matematiche e Naturali all'Accademia Nazionale dei Lincei
il 9 Marzo 2018.

Un ringraziamento particolare a Francesco Berrilli, Menico Rizzi,
Enrico Romita, Nicola Vittorio per l'attenta lettura e i preziosi
suggerimenti che hanno contribuito a migliorare il testo.

\end{document}